\documentclass[11pt]{amsart}
\usepackage{amsmath, amsfonts, amsthm, amssymb}
\usepackage[longnamesfirst]{natbib}
\usepackage{verbatim}
\usepackage{hyperref}
\usepackage{color}
\usepackage[dvips]{graphicx}
\usepackage{pst-all}
\usepackage[all]{xy}

\usepackage[margin, draft]{fixme}
\usepackage{graphicx}
\usepackage{enumerate}         
\newcommand{\Y}{\mathbf{Y}}
\newcommand{\RR}{\mathbb{R}}
\newcommand{\QQ}{\mathbb{Q}}

\newcommand{\NN}{\mathbb{N}}
\newcommand{\EE}{\mathbb{E}}

\newcommand*{\wh}{\widehat}

\begin{document}
\title[Arbitrage-free prediction of the implied volatility smile]{Arbitrage-free prediction of the implied volatility smile}

\author{Petros Dellaportas}
\address{Department of Statistics,
Athens University of Economics and Business, Athens, Greece}
\email{petros@aueb.gr}

\author{Aleksandar Mijatovi\'{c}}
\address{Department of Mathematics, Imperial College London, UK}
\email{a.mijatovic@imperial.ac.uk}


\thanks{We thank the RBS FX options desk in London for making available the USDJPY
options data used in this paper. We are grateful to Dilip Madan for suggesting
this problem to us. This work  is financed by the European Union (European Social Fund - ESF) and Greek national
funds through the Operational Program
``Education and Lifelong Learning'' of the
National Strategic Reference Framework (NSRF), project ARISTEIA-LIKEJUMPS.}






\begin{abstract}
This paper gives an arbitrage-free prediction for future prices of an arbitrary 
co-terminal set of options with a given maturity, based on the observed 
time series of these option prices. 
The statistical analysis of such a multi-dimensional time series 
of option prices corresponding to $n$ strikes (with $n$ large, e.g. $n\geq 40$) 
and the same maturity, is a difficult task due to the fact that 
option prices at any moment in time satisfy non-linear and non-explicit 
no-arbitrage restrictions. Hence any $n$-dimensional time series 
model also has to satisfy these implicit restrictions at each time step,
a condition that is impossible to meet since the model innovations can take 
arbitrary values. We solve this problem for any $n\in\NN$ in the
context of Foreign Exchange (FX) by first encoding the
option prices at each time step in terms of the parameters
of the corresponding risk-neutral measure and then performing the time series
analysis in the parameter space. The option price predictions are obtained
from the predicted risk-neutral measure by effectively integrating it
against the corresponding option payoffs.
The non-linear transformation between option prices and
the risk-neutral parameters applied here is \textit{not} arbitrary: 
it is the standard mapping used by market makers in the FX option markets
(the SABR parameterisation) and is given explicitly in closed form. 
Our method is not restricted to the FX asset class nor does it 
depend on the type of parameterisation used.
Statistical analysis of FX market data illustrates that our 
arbitrage-free predictions outperform the naive random walk 
forecasts, suggesting a potential for building management strategies
for portfolios of derivative products, akin to the ones widely used
in the underlying equity and futures markets.
\end{abstract}

\maketitle


\section{Introduction}


One might expect that the time series of past option prices or,
equivalently, their implied volatilities, could be used 
to predict their future behaviour.
However, there is a major theoretical and practical impediment 
to constructing a statistical model that can
be estimated on a time series of option prices,\footnote{Throughout the paper
we assume that a manager has access to a good quality implied volatility 
time series data. Our data set is described in Section~\ref{subsec:Desc} below.}
namely that the state space of the model has to consist of arbitrage-free option 
prices only \textit{and} the expected value of the model at any time also needs to lie 
within this state space.
Otherwise the predictions are not arbitrage-free and are hence of little use 
in practice since they can certainly not be 
fulfilled (the contemporaneously quoted option prices in the derivatives
markets are always free of arbitrage).
Since the arbitrage-free requirement is both non-linear and
non-explicit, it is impossible to satisfy it in a time series 
model for option prices, as in such a framework the model innovations are allowed to take arbitrary values 
thus making the model necessarily violate the no-arbitrage constraints with positive probability. 





This paper provides a framework that overcomes the difficulty arising from the
arbitrage-free requirement in forecasting option prices and shows that the predictions
obtained are not only internally consistent, but also outperform, 
for the data used in our empirical study, the ones based on  a model that
assumes a random walk process for the option prices. The random walk (or white
noise) model is a standard benchmark  in the related literature for forecasting
equities and it is typically very hard to beat; see, for example,
\cite{welch2008comprehensive}. It simply forecasts the option prices at time
$t+1$ by the observed price at the time  $t$ and it is by default
arbitrage-free.  We consider prediction of the 
{\em volatility smile}, which is equivalent to the prediction of option prices with
any strikes and a fixed maturity (see e.g.~\cite{BreedenLitzenberger1978, Renault_overview}
and Section~\ref{sec:risk_neutral}).  Our methodology 
consists of the following key idea. First, for a given maturity, encode the time series of the historical 
implied volatilities across strikes  in terms of the
parameters of a corresponding risk-neutral measure. Second, preform a time series
analysis on the implied historical parameter observations. We provide 
empirical evidence that, at least in the Foreign Exchange (FX) derivatives market from
which we have available data, the time series based on the observed parameter
values exhibits predictive ability for implied volatilities (and hence option
prices) through ARMA-GARCH based statistical models for the risk-neutral parameters.
The key benefit of this approach (i.e. first performing a non-linear transformation of implied volatilities
and then applying a model to the time series of the implied risk-neutral parameters) lies in the fact that, 
in the parameter space, the no-arbitrage constraints are reduced to ensuring that the state space of the 
time series model is defined by the natural parameter restrictions. 
As we shall see in Section~\ref{sec:Model}, this is typically easy to satisfy
and, in particular, implies that the forecasts also lie within the state space. 
Equivalently put, the predicted option prices will be free of arbitrage.
In Section~\ref{subsec:Trading_Strategy} a simple trading strategy for
strangles and risk-reversals, 
based on our approach and out-of-sample options data, is presented.

\section{The problem}
\label{sec:Propblem}
\subsection{Prediction of implied volatilities}

Our aim is to investigate the problem of the arbitrage-free 
forecasting of option prices in the context of the co-terminal (i.e. with the 
same expiry) call options on the USDJPY exchange rate. 
As mentioned in the introduction, 
call and put contracts on the exchange rates of the currencies of
major economies are among the most liquid 
derivatives contracts in the financial markets. 
The FX derivatives market, where such contracts trade,
is very deep, both in terms of the volume of options traded and
the size of the underlying notional of the contracts. 
We will develop our approach using call options but,
due to the put-call parity (see e.g.~\cite[Sec.~8.4]{Hull}), 
our results are directly applicable to put options.

Recall that 
a call option struck at 
$K$
and expiring 
at some future time 
$T$
is a contract that gives the owner the right,
but not the obligation, to purchase 
$N\$ $
at expiry
$T$
for the price of 
$K$ $\yen$
per $\$$.
Furthermore, the price $C(K,T)$
of a vanilla call option described above is well-defined 
for practical purposes,
because the bid-offer spread is tight
due to the volume, frequency and size of the trades.
The market makers in FX options keep a 
close eye on and record throughout a trading day
the prices of all the options 
on a given currency pair across strikes and maturities.
In practice, option traders do not refer to option prices 
\textit{per se} but rather express them through 
the Implied Volatility (IV) metric, which allows 
one to compare directly option prices across different strikes,
maturities, expiration times and underlyings
in the same units
(IV is derived using the Black-Scholes formula, see Section~\ref{sec:IVol} 
for definition and~\cite{Gatheral} for more details).
For this reason, 
in what follows we may (and shall) consider the prediction of the 
implied volatilities
$IV(K,T)$ 
instead of the option prices 
$C(K,T)$,
without loss of generality and with the benefits mentioned above.

The main obstacle to a direct application of econometric tools to the
multi-dimensional time series of option prices can be described as follows.  To
make the discussion simpler let us fix a maturity $T$ (e.g. 1 month).  Note
first that there are uncountably many possible option contracts with maturity
$T$ at any moment in time as there are uncountably many possible strikes  $K$.
Assume further that we have chosen strikes $K_i$, $i=1,\ldots,n$ (with e.g.
$n=40$), and would like to model the evolution of the corresponding implied
volatility prices $IV_t(K_i,T)$, $i=1,\ldots,n$, over time (indexed by
$t\in\NN$).  Let the $n$-dimensional vector $\Y_t$, with co-ordinates given by
$$ \Y_t(i):=IV_t(K_i,T),\qquad i=1,\ldots,n, $$ 
describe the midday implied
volatilities on USDJPY on day $t$ for all the strikes $K_i$ and assume that we
are given such implied volatilities over the time period $t=1,\ldots,t_0$.  A
direct approach based on a general multi-dimensional time series model predicts
$\Y_{t+1}$  with a predictor $G(\Y_{1:t})$ based on a general function $G$, by
minimizing the squared error loss $E(\Y_{t+1} - G(\Y_{1:t}))^2$, or,
equivalently, by setting $G(\Y_{1:t})$  as the conditional mean $E(\Y_{t+1} |
\Y_{1:t})$.  Since the above minimum mean-squared error predictor of $\Y_{t+1}$
given $\Y_{1:t}$ presupposes knowledge of the joint distribution of $\Y_{1:t+1}$,
which is not available, we assume that it belongs to a parametric family of
distributions indexed by a parameter vector $\theta$.  Then, the optimality of
the above predictor cannot be obtained since for every $\theta$ we shall have a
different optimal predictor $E(\Y_{t+1} | \Y_{1:t}, \theta)$.  However, a
simple way to proceed that gives reasonable predictions is to replace the
unknown parameter $\theta$ with a suitable estimate based on past history
$\Y_{1:t}$.  What makes this approach infeasible to our problem of predicting
implied volatilities is the fact that at the time step $t+1$, the components of
the vector $\Y_{t+1}$, i.e. the modelled implied volatilities
$IV_{t+1}(K_i,T)$, $i=1,\ldots,n$, have to satisfy non-linear and non-explicit
no-arbitrage restrictions.
It is an extremely hard problem to identify an
estimate of $\theta$ that will result in a state vector $\Y_{t+1}$ with the
desired no-arbitrage restrictions.\footnote{It follows 
from~\cite{BreedenLitzenberger1978}  
that, when translated back to the call option prices 
$C(K,T)$,
the no-arbitrage restrictions amount to 
the function 
$K\mapsto C(K,T)$
being non-increasing and convex. However, imposing such restrictions 
on the stochastic dynamics of 
$C_{t}(K_i,T)$, $i=1,\ldots,n$, 
at each moment of the running time 
$t$
constitutes a problem of equivalent complexity.}

Before describing the solution to this problem,
we should stress here that, due to the
liquidity in the FX options market, the recorded
implied volatilities always satisfy the no-arbitrage 
requirements, and therefore it is hard to perform the statistical 
analysis of the options data set. This fact by no means implies that in 
a non-efficient market the statistical analysis of
implied volatilities would be easy. On the contrary, if 
the options market is not efficient
enough to have contemporaneous implied volatility quotes
across a family of strikes for a given maturity, 
the problem is not well-defined because there is
no internally consistent time series of implied volatilities
and a more sophisticated approach for the estimation of the
pricing kernel (see e.g.~\cite{XMM}) is required before any kind
of statistical analysis can begin.

\subsection{A solution}
The above discussion indicates that the arbitrage-free prediction 
of implied volatilities is a difficult problem, 
a solution to which has, to our knowledge, not yet been proposed. 
We propose a solution below, which will be shown consistently
to outperform random walk predictions in a large out-of-sample
exercise (see Figure~\ref{jpyiv} in Section~\ref{subsec:Time_Seris_Modelling}
for the numerical results).

In order to predict the implied volatility vector
$\Y_{t+1}$,
we first apply a non-linear transformation 
to the implied volatilities data up to and including time
$t$.
This one-to-one transformation maps the implied volatilities into
the parameters of a risk-neutral measure 
$\QQ$
(see Section~\ref{sec:risk_neutral}
for the definition and properties of the
parametric form of the risk-neutral measure used 
in FX options markets).
The core of our approach rests on the fact that 
every element in the space of $\QQ$
parameters yields an arbitrage-free set of implied volatilities
across all strikes.
This allows us to perform the statistical analysis 
in the space of $\QQ$-parameters.
We use the daily recorded parameters 
that characterise the choice of the risk-neutral measure
given by the quoted implied volatilities on that day,
to forecast, via a 
time series model, 
tomorrow's $\QQ$-parameter values
(see Section~\ref{subsec:Desc} for the precise
description of our data set).
The predicted implied volatilities
$\wh \Y_{t+1}$
are then computed via a non-linear pricing function 
that guarantees that
the no-arbitrage restrictions are satisfied. 
A schematic description of our approach to this
problem is given in the following diagram:
$$
\begin{array}{c@{\hspace{3cm}}c}
\Rnode{N1}{\text{IV data:}\> 1:t}
&\Rnode{N2}{\text{Prediction:}\quad \wh \Y_{t+1}
}\\[2cm]
\Rnode{N3}{\text{$\QQ$-parameters}} 
&\Rnode{N4}{\text{Predicted $\QQ$-parameters}} 
\end{array}                                
\psset{nodesep=0.3cm}
\everypsbox{\scriptstyle}
\ncLine{->}{N1}{N3}\Bput{\text{Inverse pricing formula}}
\ncLine{->}{N3}{N4}\Aput{\hspace{-10mm}\text{Stochastic model}}
\ncLine{->}{N4}{N2}\Aput{\text{Pricing formula}}
$$

In the mathematical finance literature, arbitrage-free evolution 
of option prices under a risk-neutral measure has been studied extensively (see 
e.g.~\cite{Carmona, Kallsen, Schweizer} and the references therein).
While these approaches are both mathematically involved and very interesting,
we should point out that the problem of the arbitrage-free evolution 
of the implied volatility smile/surface under the risk-neutral measure
is distinct from the main aim of the present paper. In this paper we 
model the evolution of the implied volatility smile under the real-world measure and are
solely concerned with the arbitrage-free constraint in a static sense,
i.e. at each moment of the running time,\footnote{This is completely
analogous to the way pricing theory is applied in the derivatives markets.}
in order to obtain 
viable predictions for tomorrow's 
option prices. 
Furthermore, note that the stochastic dynamics under the 
continuous-time models, like the ones in~\cite{Carmona, Kallsen, Schweizer},
is typically very complicated making it unclear 
how and whether such models can 
be efficiently estimated and applied for the statistical prediction of 
future option prices.

\section{Arbitrage-free option prices in Foreign Exchange} 
\label{sec:risk_neutral}

Intuitively, 
a set of implied volatilities
$IV_t(K_i,T)$, 
$i=1,\ldots,n$,
allows arbitrage
if and only if 
it is possible to trade (i.e. buy and sell)
the corresponding contracts, 
at prices given by
$IV_t(K_i,T)$, 
$i=1,\ldots,n$,
and follow a self-financing trading strategy
on the FX spot rate
(i.e. buy and sell the underlying currencies) 
in such a way that at some future time the 
portfolio of options and the gains (i.e.
P\&L) from the trading strategy 
in agregat have a non-negative value with certainty 
and a strictly positive value with positive probability. 
In other words, if a market maker (i.e. a large trading
desk whose business it is to buy or sell options of any
strike and maturity) quotes option prices that
allow arbitrage, a counterparty could enter into a 
contract with the market maker  
{\em without any risk of loss}
(here we are allowed to ignore the fact that 
prices are quoted in terms of bids and offers,
because, as mentioned in Section~\ref{sec:Propblem},
the spread in USDJPY options is typically 
only a few basis points).
It is evident that, when the number of option prices is large
(it is in the thousands or even tens of thousands, in the case 
of market makers) it is very hard to check \textit{a priori}
from the quoted option prices whether they are arbitrage-free.
In practice this is achieved by using the theoretical concept 
of the \textit{risk-neutral} measure, which arises
in the fundamental theorem of asset pricing: 
the implied volatility prices 
$IV_t(K_i,T)$, $i=1,\ldots,n$,
satisfy a no-arbitrage restriction if they are obtained 
as discounted expectations of their respective payoffs
under a risk-neutral measure
$\QQ$. In other words, if they are of the form
$$
e^{-(r_d-r_f)T}\EE^\QQ\left[(S_T-K_i)^+\right],
$$
where
$S_T$
denotes the FX rate (e.g. USDJPY) at the future time 
$T$,
$r_d$
(resp.
$r_f$)
represent 
the dollar (resp. yen)
interest rate over the time interval of length 
$T$
and,
$x^+=\max\{0,x\}$
for any
$x\in\RR$.
The defining feature of a risk-neutral measure
$\QQ$
is that the mean of the random 
FX rate 
$S_T$
at time 
$T$
is equal to the exponential of the interest rate differentials
of the two currencies over the time period from now until
expiry
$T$:
\begin{equation}
\label{eq:Risk_neutral_drift}
\EE^\QQ[S_T] = F_T,\qquad\text{where}\qquad
F_T=e^{(r_d-r_f)T}S_0.
\end{equation}

The condition in~\eqref{eq:Risk_neutral_drift}
is clearly satisfied by many probability measures 
$\QQ$,
as it only fixes the mean of the random variable 
$S_T$.
This gives the market makers
(and other market participants) a large amount of freedom in choosing 
the option prices in a no-arbitrage way as there clearly exists a vast
family of probability measures, and hence risk-neutral models
for the FX rate $S_T$,
such that~\eqref{eq:Risk_neutral_drift}
holds. 

The following two observations play an important role in how the concept of the
risk-neutral measure is applied in the financial markets: (a) The choice of a risk-neutral
measure does not necessarily give the correct (i.e. market) price for a given
derivative contract. All such a choice does is to provide arbitrage-free prices for
all traded options with the same maturity. Note that, since the market maker is
interested in the consistent pricing of derivatives, this is precisely what they are
after. (b) The correct price of an option, or equivalently the correct choice of the
risk-neutral measure, is down to supply and demand in the market and crucially
depends on the market maker’s view of where the specific contracts should trade.
This is why flexible parametric forms of the risk-neutral measure are useful: they
allow the market maker to express their view of where the liquid derivatives should
trade, while providing an arbitrage-free pricing mechanism for all derivatives.


As it turns out, in the FX 
option markets there is 
a standard parametric form for the choice of the risk-neutral measure,
which we will describe in detail in Sections~\ref{sec:IVol} and~\ref{sec:SABR}.
This not only allows us to circumvent the difficulty arising from the no-arbitrage 
restriction for option prices, but it expresses the vanilla option for any strike 
and a given maturity in terms of the values of three parameters, the current level
of the FX spot rate and the two interest rates in the respective currencies 
over the period from current time until maturity. 
There are two reasons why the particular parameterisation of option prices,
based on the SABR formula (see Section~\ref{sec:SABR}) has
been adopted as market standard:
\begin{enumerate}[(i)]
\item the parsimonious description of all option prices with a given maturity 
is very robust and easy to use for the traders;
\item each of the three parameter values is closely related to the
three fundamental features of the risk-neutral distribution (which determines 
option prices): the over-all level of option prices and 
the skweness and kurtosis of 
the risk-neutral law 
of the FX spot rate $S_T$.
\end{enumerate}

It should be stressed here that 
the analytical SABR formula 
(see Equation~\eqref{eq:SABR} below)
in general yields arbitrage-free 
option prices except in the extreme wings
(i.e. for the strikes that are many standard deviations
away from the at-the-money value).
However, the points made above and the fact that
for such extreme strikes the notion of the market defined
price is questionable due to the lack of liquidity,\footnote{We thank
the referee for this observation.}
suggest that it is as reasonable to apply the SABR parametrisation of the 
risk-neutral measure for purposes of our problem as it is to use
it for the pricing, hedging and risk management of
the portfolios of options, a very common day-to-day 
practice in the FX options markets. A further argument in favour of using
the SABR formula, besides the lack of liquidity in the wings, is
that the price of options in the extreme wings is more uncertain than 
in the ``near'' wings due to the 
increase in the bid-offer spread for such derivatives.

\subsection{Implied volatility in the FX markets}
\label{sec:IVol}

The value
$\textrm{BS}(F_T,K,T,\sigma)$
of the European call option with strike
$K$
and expiry
$T$
in
a Black-Scholes
model
with constant volatility
$\sigma>0$
is given by
the Black-Scholes formula
\begin{eqnarray}
\label{eq:BS_Formula}
\textrm{BS}(F_T,K,T,\sigma) & := & e^{-r_dT}\left[F_T \,N(d_+)-K\,N(d_-)\right],\qquad
\end{eqnarray}
where
\begin{eqnarray*}
d_{\pm}=\frac{\log(F_T/K)\pm\sigma^2T/2}{\sigma\sqrt{T}},
\end{eqnarray*}
and
$N(\cdot)$
is the standard normal cumulative distribution function.
The implied volatility
that corresponds to the 
market price
$C(K,T)$
for the strike
$K>0$
and
maturity
$T>0$
is the unique positive number
$IV(K,T)$
that satisfies the following equation in the variable
$\sigma$:
\begin{eqnarray}
\label{eq:DefOfImpiedVol}
\textrm{BS}\left(F_T,K,T,\sigma\right)=C(K,T).
\end{eqnarray}
Implied volatility is well-defined since
the function
$\sigma\mapsto \textrm{BS}\left(F_T,K,T,\sigma\right)$
is strictly increasing for
positive
$\sigma$
(the \textit{vega} of a call option
$\frac{\partial \textrm{BS}}{\partial \sigma}(F_T,K,T,\sigma)
=e^{-r_dT}F_T N'(d_+)\sqrt{T}$
is clearly strictly positive) and
the right-hand side of~\eqref{eq:DefOfImpiedVol}
lies in the image of the Black-Scholes formula
if and only if the call option price
$C(K,T)$
satisfies a no-arbitrage restriction.
It is clear from the definition,
albeit suppressed in the notation, that the implied volatility
$IV(K,T)$
also depends on the current level of the FX spot rate
$S_0$
and the interest rate differential between the two currencies.

As noted earlier, the implied volatility
$IV(K,T)$
is nothing more than a convenient number (in $\%$)
to express the price of a call option with expiry 
$T$
and
strike
$K$
via the Black-Scholes formula in~\eqref{eq:BS_Formula}.
The convenience, from the perspective of market participants,
lies in the fact that 
$IV(K,T)$
contains information about the price of a call option, which 
is independent of the FX rate and can therefore be easily compared
across currency pairs. 
Furthermore, implied volatility is being used in the FX option markets 
to specify parametrically the call option prices for all strikes and a given
maturity. It is well-known (see e.g.~\cite{BreedenLitzenberger1978}) that 
the information contained in knowing the prices of the implied volatilities 
for all strikes and a given maturity $T$ is equivalent to specifying 
the risk-neutral law of the FX rate 
$S_T$.
Therefore, the SABR formula for the implied volatility
$IV(K,T)$
given in the next section together with the Black-Scholes
formula in~\eqref{eq:BS_Formula}
specify an easy parameterisation of the risk-neutral law
of the spot rate
$S_T$.

\subsection{The parametric form of the risk-neutral measure in the FX markets}
\label{sec:SABR}

The version of the SABR formula for the implied volatility, derived in~\cite{SABR2002},
which will be used in this paper 
takes the form
\begin{eqnarray}
\label{eq:SABR}
IV(K,T) & = & 
\alpha\frac{1+T\left[\frac{1}{4\cdot24}\frac{\alpha^2}{(F_TK)^{1/2}}+\frac{1}{4}\frac{\rho\nu\alpha/2}{(F_TK)^{1/4}}+\frac{2-3\rho^2}{24}\nu^2\right]}
{(F_TK)^{1/4}\left\{1+\frac{1}{4\cdot24}\log^2\left(\frac{F_T}{K}\right)+\frac{1}{16\cdot1920}\log^4\left(\frac{F_T}{K}\right)\right\}}
\cdot \frac{z}{x(z)}
\end{eqnarray}
where 
\begin{eqnarray*}
z  =  \frac{\nu}{\alpha} (F_TK)^{1/4} \log\left(\frac{F_T}{K}\right) &\text{and}&
x(z)  =  \log\left\{\frac{\sqrt{1-2\rho z +z^2}+z-\rho}{1-\rho}\right\}.
\end{eqnarray*}
The market data in this formula consist of the current FX  spot rate
$S_0$ and the interest rate differential for the maturity 
$T$, which features in 
$F_T$
(see formula in~\eqref{eq:Risk_neutral_drift}).
The parameters that allow the market participant to express their view on the price
of the call option struck at 
$K$
with maturity
$T$
are given by
\begin{eqnarray*}
\alpha & & \qquad\text{instantaneous (current) level of volatility,}\\
\nu & & \qquad\text{volatility of volatility,}\\
\rho & & \qquad\text{instantaneous correlation.}
\end{eqnarray*}
A general version of the SABR formula contains
a further parameter $\beta$ the value of which 
is taken to be 
$1/2$ 
in~\eqref{eq:SABR}.
In practice 
$\beta=1$
is also used, marginally changing the analytical expression 
in~\eqref{eq:SABR}. However, this makes no material difference
for pricing purposes as a change in $\beta$
can be compensated by a change 
in the value of 
$\rho$
since both parameters mainly effect the skew of the 
smile (cf. the final paragraph of this subsection).\footnote{We thank one of the referees for this remark.}

The formula in~\eqref{eq:SABR}
yields a natural parameterisation of the risk-neutral
law of 
$S_T$.
It was developed in~\cite{SABR2002}
based on an assumption that the FX spot rate process
$(S_t)_{t\geq0}$
evolves under a risk-neutral measure as a stochastic volatility
process: 
$\alpha$
denotes the value of the volatility process in this model
at time zero (i.e. the current value), 
$\rho$
is the correlation between the two Brownian motions driving the 
spot and the volatility processes, 
and 
$\nu$
is the volatility of the stochastic volatility process; 
see~\cite{SABR2002} for more details.

A simple and yet important fact is that for each value of the 
state-vector 
$(\alpha,\nu,\rho)$,
where
$\alpha$ and $\nu$
are positive 
and
$\rho$ is between $-1$ and $1$,
the function
$K\mapsto \textrm{BS}(F_T,K,T,IV(K,T))$,
given by the formulae~\eqref{eq:BS_Formula}
and~\eqref{eq:SABR},
represents an arbitrage-free collection of option prices
for any positive strike
$K$.
The parameters 
$(\alpha,\nu,\rho)$
are used to control the risk-neutral distribution in the following
way: a change in the parameter $\alpha$
has the effect of changing the overall level of option prices,
the parameter
$\nu$
and
$\rho$
control the kurtosis 
and skewness of the risk-neutral distribution 
of
$S_T$.
This natural interpretation of the parameters has made~\eqref{eq:SABR}
a standard parameterisation of a risk-neutral law for the spot
in the FX option markets. 

It should be stressed, however, that our approach does not depend
on the specific parametric form of a risk-neutral law. We use
the time series of parameters of formula~\eqref{eq:SABR}, because
they describe the market implied risk-neutral law of the FX rate
$S_T$
extremely accurately: since the formula has emerged as  the market
standard, the values of the parameters 
$(\alpha,\nu,\rho)$
are used by the traders to express the market consensus on what
shape the risk-neutral law of 
$S_T$
takes. 
In particular, the parameter $\beta$
mentioned in the first paragraph of this subsection
could have taken the value $\beta=1$
(instead of $\beta=1/2$)
without ramifications for our approach.
More generally, 
in a different market the method outlined here can be applied to the parameters
of the risk-neutral law  given by the standard
of that particular market. For example in the equity 
option markets one can use Heston's parameterisation~\cite{Heston} of the 
risk-neutral law.

\section{Arbitrage-free modelling of co-terminal options in Foreign Exchange}
\label{sec:Model}

\subsection{Description of the data set and FX option market conventions}
\label{subsec:Desc}
In foreign exchange the liquid options of a given market-defined maturity
$T$,
e.g. 
$T=\text{1 month}$,
have a rolling expiry with respect to the current time $t$. 
In other words, at time $t$
a liquid call option expiring  in one month would
cover the time interval
$[t,t+T]$.
The next day, i.e. 
at time 
$t+1$,
a liquid one-month option will be a different derivative
contract covering the time interval
$[t+1,t+1+T]$.

Our data set consists of the time series for the parameter values
$(\alpha, \nu,\rho)$
implied by the  USDJPY liquid one-month options,
the FX spot rate level and the interest rate differentials 
at London midday over the period 29/9/2006 to 16/12/2011.
The data set has been provided by the RBS FX options desk,
one of the largest market makers in FX options.
The option quotes are taken at London midday to ensure that 
all the option prices are temporally consistent. In other words, 
if the option prices that are used to obtain the parameters
$(\alpha, \nu,\rho)$
are not recorded simultaneously, there is no guarantee that 
the parameter triplet reflects the market view on the risk-neutral
law of the FX spot rate.
Since London midday is the only time at which the trading desk 
records all the liquid option prices simultaneously, 
the parameters
$(\alpha, \nu,\rho)$
obtained in this way describe all the option prices with expiry 
$T$
without the usual difficulty with options data arising from
the illiquidity in the market. Furthermore, this also gives us
a very clear interpretation for the predicted parameters:
our prediction yields a risk-neutral law of 
$S_T$
that can be converted to a vector of option prices 
and compared (out-of-sample) to the option prices 
that are observed at future London midday fixing times.

\subsection{Time series modelling}
\label{subsec:Time_Seris_Modelling}

We have obtained the data set
$(\alpha_t, \nu_t,\rho_t)$ for $t=1,\ldots,1360$ 
for the USDJPY exchange rate. Our modelling 
construction is executed as follows:
we use the first $n_0=1000$ values to
fit a model and the next $n_1 = 360$ values 
to test its predictive ability.
We first transform our data using 
\begin{eqnarray}
A_t & := & \log ( \alpha_t )   \nonumber \\
N_t & := & \log (\nu_t)  \nonumber \\
R_t & := &\log \left( \frac{\rho_t+1}{1-\rho_t} \right) \nonumber
\end{eqnarray}
so that all transformed parameters lie in the real line.  This will guarantee that the predictions achieved through the time series models obey the restrictions $\alpha>0, \nu>0$ and $-1<\rho<1$. 
Model building has been guided through an ARMA-GARCH time-series methodology
which includes autocorrelation and partial autocorrelation plots, AIC and BIC
information criteria and parameter significance tests;  see, for example,
\cite{brockwell2005time}.  In particular, in all series  there was overwhelming
evidence that the hypothesis of unit root cannot be rejected (with augmented
Dickey-Fuller and Phillips-Peron tests) so the difference operator was applied
to all series and in the resulting series the unit root hypothesis was
rejected.  All (differenced) series exhibited strong GARCH effects that were
evident by both inspecting the autocorrelation plots of their squares and by
Engle's ARCH test.  After removing the GARCH effects (in which student-t errors
seemed to perform much better than Normal errors when BIC and AIC values were
compared and when residuals were inspected) the autocorrelation and partial
autocorrelation plots of the residual series indicated that forms of ARMA-GARCH
models might be appropriate.  Our subsequent modelling choice procedure
consisted of fitting a series of such models  and finally the `best' models
have been chosen with respect to the best AIC/BIC values attained.  In some
ARMA-GARCH models the problem of near root cancellation that results in
misleading inference (see, for example, \cite{ansley1980finite}) was observed
so we chose to remove the moving average component.  Parameter estimates have
been obtained through MATLAB software. A mathematical formulation of our models
is as follows.

By first defining, for $t=2,\ldots,n_0$,  $\Delta \alpha_t  :=  A_t - A_{t-1} $,
$\Delta \nu_t  :=  N_t - N_{t-1} $, and $\Delta \rho_t  :=  R_t - R_{t-1} $,
 we obtained as best models the following general
specifications:
\begin{eqnarray}
\Delta \alpha_t & = & 
\mu^\alpha + 
\sum_{i=1}^{p^\alpha} \phi^\alpha_i \Delta \alpha_{t-i} +
\sum_{j=1}^{q^\alpha} \theta^\alpha_j \epsilon^\alpha_{t-j} + 
\epsilon^\alpha_t 
\nonumber \\
\Delta \nu_t & = & 
\mu^\nu + 
\sum_{i=1}^{p^\nu} \phi^\nu_i \Delta \nu_{t-i} +
\sum_{j=1}^{q^\nu} \theta^\nu_j \epsilon^\nu_{t-j} + 
\epsilon^\nu_t 
\nonumber \\
 \Delta \rho_t & = & 
\mu^\rho + 
\sum_{i=1}^{p^\rho} \phi^\rho_i \Delta \rho_{t-i} +
\sum_{j=1}^{q^\rho} \theta^\rho_j \epsilon^\rho_{t-j} + 
\epsilon^\rho_t 
\nonumber 
\end{eqnarray}
where the error terms $\epsilon_t^\alpha,\epsilon_t^\nu,\epsilon_t^\rho$ follow
a GARCH(1,1) model with Student-t errors. 
The resulting estimated parameters,
based on the initial time series of $1000$ data points, are given in 
Table~\ref{table:parameters}.

\begin{table}[ht!]
\renewcommand{\arraystretch}{0.7}
  \begin{center}
    \begin{tabular}{|c||c|c|c|c|c|}
      \hline
     \multicolumn{6}{|c|}{\bf Parameter estimates for the USDJPY series} \\
      \hline
$\mu^\alpha$ & -0.0002 &&&&\\ 
  & (0.0001) &&&&\\ 
 $\phi^\alpha$ & 0.9104 & &  &  &\\ 
  & (0.0160) & &  &  &\\ 
 $\theta^\alpha$ & -0.9789 &  &  & &\\ 
  & (0.0053) &  &  &  &\\ 
 $\epsilon^\sigma$ & 0.0002 & 0.1801 & 0.7807 & 3.8903 &\\ 
  & (0.0001) & (0.0444) & (0.0413) & (0.5788) &\\ 
  \hline
$\phi^\nu$ & -0.1844& -0.2279 & -0.2269 & -0.1096 & 0.2753   \\ 
  & (0.1082) & (0.0372) & (0.0441) & (0.0428) & (0.0341)    \\ 
 $\theta^\nu$ & 0.0042  &&&&\\ 
  & (0.1105)  &&&&\\ 
 $\epsilon^\nu$ & 0.0000 & 0.0317 & 0.9397 & 6.1987 &\\ 
  & (0.0000) & (0.0145) & (0.0311) & (1.3818) &\\ 
		\hline
	 $\mu^\rho$ & 0.0013 &&&&\\ 
   & (0.0022) &&&&\\ 
 $\phi^\rho$ & -0.0520  &&&&\\ 
  & (0.0273)  &&&&\\ 
 $\epsilon^\rho$ & 0.0004 & 0.0445 & 0.9262 & 2.9694 &\\ 
  & (0.0003) & (0.0249) & (0.0444) & (0.4177) &\\  \hline  
	\end{tabular}
  \end{center}

\vspace{10mm}

\caption{Estimated parameters (standard errors in brackets) for the 
time series models.  The four $\epsilon$ parameteres correspond to the GARCH(1,1)
model estimates: the constant, the ARCH parameter, the GARCH parameter and the
degrees of freedom of the t-density respectively.}
\label{table:parameters}
\end{table}

\begin{figure}[h]
\centerline{\includegraphics[width=4in]{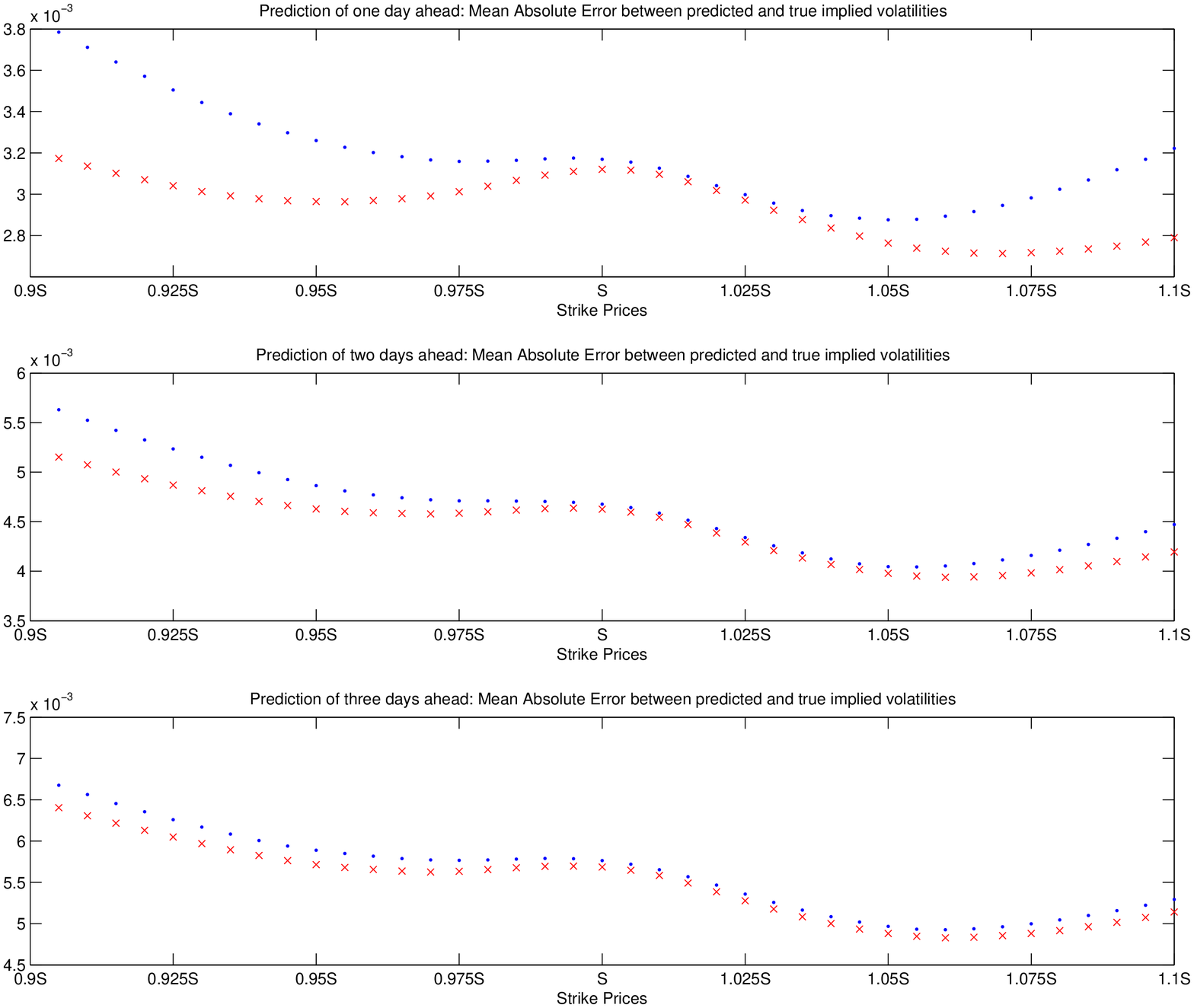}}
\caption{USDJPY. Red cross: our predictions; blue dots: random walk predictions.}
\label{jpyiv}
\end{figure}

The out-of-sample prediction exercise was performed by fitting the model of
Table~\ref{table:parameters} as each new data point $t=1001,1002,\ldots,1360$ arrives
and then predicting the triplet of the parameters through the fitted model for
one, two and three days ahead.  These predicted triplets were then transformed
back to predicted implied volatilities via~\eqref{eq:SABR}.  The three
ingredients that were unknown in~\eqref{eq:SABR}, namely the future spot  
rate $S_0$
and the two interest rates $r_d$, $r_f$
(yielding the future forward rate
$F_T$,
see~\eqref{eq:Risk_neutral_drift}),
have been predicted with a simple random walk
model, so were taken to be equal to current values.  Strikes $K$ were
selected to be forty equally spaced values between $90\%$ and $110\%$ of
the current spot rate. For every future day and every strike we have
calculated the predicted implied volatility and recorded the absolute error
against the implied volatility derived using the true values of the parameters
$\alpha, \nu,\rho$ and (of course)
the realised values of the spot rate
$S_0$
and the interest rates
$r_d,r_f$.  
Figure~\ref{jpyiv} illustrates that the
mean error varies between $28$ basis points for predictions 
one day ahead and may even reach $65$ basis points for 
predictions three days ahead. 

In Figure~\ref{jpyiv} we have depicted (with blue dots) the corresponding errors
produced by the naive random walk predictor (i.e. tomorrow's option prices
are the same as today's). Our methodology clearly
outperforms the random walk forecasts for all strikes
with the improvement reaching up to four basis points 
for out-of-the-money options.  Beyond the three day
time horizon, this improvement vanishes, indicating
that our ARMA-GARCH model does not have further 
predictive ability.

An important observation is that the predicted risk-neutral measure
in our framework depends, not only on the SABR parameters, which 
our methodology predicts, but also on the future level of the
FX forward rate. Since the latter is notoriously difficult to 
predict, in our approach we use the random walk prediction for it.
Recall that the level of the forward rate is given by the appropriately 
scaled mean of the FX rate under the risk-neutral measure,
see~\eqref{eq:Risk_neutral_drift}. 
Figure~\ref{jpyiv}
clearly shows that our forecasts outperform 
the random walk by a larger amount for options that are
struck further out-of-the-money (i.e. their strike is far from
the FX forward rate). The shape of the payoff of such options 
dictates that their price depends to a greater extent on the tail 
and less so on the mean of the predicted risk-neutral distribution. 
Furthermore, it is interesting to note that,
as clearly shown by Figure~\ref{jpyiv},
historical option price data contain sufficient information 
to improve the prediction of future option prices compared to 
a random walk forecast even though the future option prices depend
on the future FX forward rate, which is very hard to forecast.
Perhaps this offers an avenue for the improved prediction of 
the FX rate itself.





\subsection{Simple trading strategy for strangles and risk-reversals}
\label{subsec:Trading_Strategy}
Based on the results of the previous section, we perform an
illustrative trading exercise to demonstrate the potential of our methodology.
We choose to buy or sell strangles ($P(K_-,T)+C(K_+,T)$)
and risk-reversals ($C(K_+,T)-P(K_-,T)$)
at
strikes
$K_-=0.9S$
and
$K_+=1.1S$,
where
$S$
is the current level of spot.
We emphasize here that this choice of
$K_\pm$
is arbitrary and that structures consisting of any finite combinations of
options with maturity $T$ may be traded, since we have arbitrage-free forecasts
for the entire implied volatility smile at $T$.
A realistic strategy requires a trading rule in which trades are executed only
when the return forecast of a strangle/risk-reversal
exceeds a given value of $\delta>0$.
Put differently 
we go long a strangle if the current 
($P(K_-,T)$, 
$C(K_+,T)$)
and predicted 
($\widetilde P(K_-,T)$,
$\widetilde C(K_+,T)$)
prices satisfy
$$
\min\left\{\frac{\widetilde C(K_+,T)-C(K_+,T)}{C(K_+,T)},
\frac{\widetilde P(K_-,T)-P(K_-,T)}{P(K_-,T)}\right\}>\delta,
$$
and act analogously in the case of a risk-reversal.
The realised P\&L of the trade is the next day's price of the structure
minus
$P(K_-,T)+C(K_+,T)$
(if we were long a strangle on the previous day).

\begin{figure}[h]
\centerline{\includegraphics[width=4in]{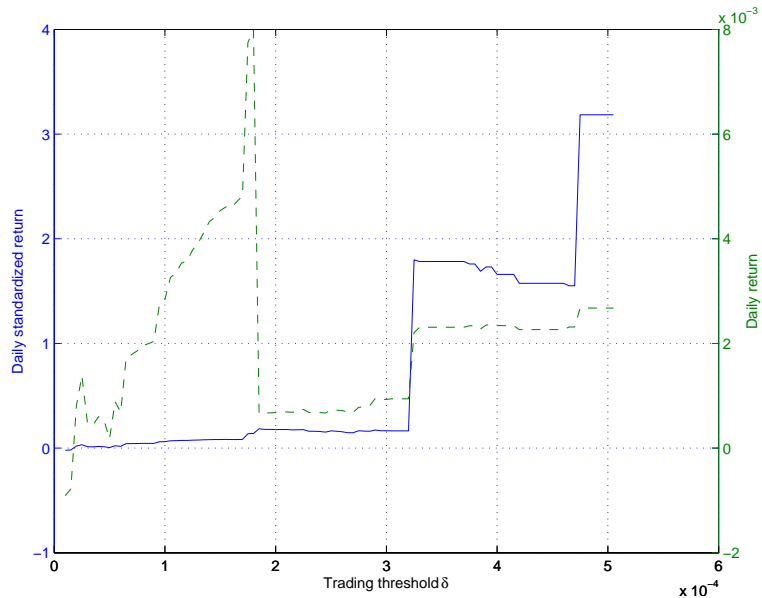}}
\caption{Performance of an out-of-sample  trading strategy.
Dashed green line: average daily return;
Solid blue line: average daily standardized return, i.e.
average daily return divided by the standard deviation of the return.
It should be noted that for
$\delta\sim10^{-5}$,
the strategy trades on $82\%$ of days (i.e. 281 out of 342)
and for
$\delta\sim 5*10^{-4}$,
it trades on 
$3.2\%$ of days (i.e. 11 out of 342).
We could of course increase the trading frequency 
for any fixed threshold 
$\delta$
by exploiting all possible structures of vanilla options
(expiring at $T$) to
obtain a stronger signal. This is possible because we have
at our disposal all arbitrage-free option prices for maturity $T$.}
\label{trading}
\end{figure}

Figure~\ref{trading} depicts the out-of-sample performance of this strategy for
the data used in Section~\ref{subsec:Time_Seris_Modelling}.  The results indicate that there is consistent
positive return expressed by both the average daily standardized return and the
average daily return of the strategy.  
Notwithstanding that 
the P\&L reported in Figure~\ref{trading} 
is gross of transaction costs, 
we believe that there is potential to adopt the arbitrage-free option price 
forecasting methodology described in this paper both for risk
management purposes and as a profitable quantitative trading strategy.
It is worth emphasising in this connection that both the trading strategy 
and the forecasting model we have used were chosen for their simplicity,
rather than their ability to capture the full potential of the approach.

\bibliographystyle{plainnat}
\bibliography{references}

\end{document}